\documentclass[12]{article}
\usepackage{fleqn}
\usepackage{graphicx}
\usepackage{amssymb}
\begin{document}
\Large
\begin{center}
{\bf The Veldkamp Space of the Smallest Slim Dense\\ Near Hexagon}
\end{center}
\large
\vspace*{.2cm}
\begin{center}
Richard M. Green$^{1}$ and Metod Saniga$^{2}$
\end{center}
\vspace*{-.5cm} \normalsize
\begin{center}
$^{1}$Department of Mathematics, University of Colorado, Campus
Box 395\\ Boulder CO 80309-0395,
U. S. A. \\
(rmg@euclid.colorado.edu)

\vspace{0.3cm}

$^{2}$Astronomical Institute, Slovak Academy of Sciences\\
SK-05960 Tatransk\' a Lomnica, Slovak Republic\\
(msaniga@astro.sk)

\vspace*{.5cm}

(6 July 2012)

\end{center}

\vspace*{.0cm} \noindent \hrulefill

\vspace*{.0cm} \noindent {\bf Abstract}

\noindent We give a detailed description of the Veldkamp space of
the smallest slim dense near hexagon. This space is isomorphic to
PG$(7, 2)$ and its $2^8 - 1 = 255$ Veldkamp points (that is,
geometric hyperplanes of the near hexagon) fall into five distinct
classes, each of which is uniquely characterized by the number of
points/lines as well as by a sequence of the cardinalities of
points of given orders and/or  that of (grid-)quads of given
types. For each type we also give its weight,  stabilizer group
within the full automorphism group of the near hexagon and the
total number of copies. The totality of (255 choose 2)/3 = 10\,795
Veldkamp lines split into 41 different types. We give a complete
classification of them in terms of the properties of their cores
(i.\,e., subconfigurations of points and lines common to all the
three hyperplanes comprising a given Veldkamp line) and the types
of the hyperplanes they are composed of. These findings may lend
themselves into important physical applications, especially in
view of recent emergence of a variety of closely related finite
geometrical concepts linking quantum information with black holes.

\vspace*{-.2cm} \noindent \hrulefill

\vspace*{.1cm}
%\large
\section{Introduction}
It has only  relatively recently been recognized that the
properties of certain finite groups and the structure of certain
finite geometries/point-line incidence structures are tied very
closely to each other. There exists, in particular, a large family
of groups relevant for physics where the (non)commutativity of two
distinct elements can be expressed in the language of finite
symplectic polar spaces (the corresponding points (not) being
joined by an isotropic line; see, e.\,g., \cite{hup}--\cite{hos})
and/or finite projective ring lines (the corresponding unimodular
vectors (not) lying on the same free cyclic submodule; see,
e.\,g., \cite{psk}--\cite{alb}). Most recently, invoking also some
finite generalized polygons, this link has been employed in
\cite{lsv,lsvp} to shed a novel light on and get deeper insights
into the so-called black hole analogy --- a still puzzling formal
relation between the entropy of certain stringy black holes and
the entanglement properties of some small-level quantum systems
(for a recent comprehensive review, see \cite{borsten}). A concept
that played a crucial role in the latter discovery turned out to
be that of the Veldkamp space of a point-line incidence structure
\cite{sglpv,vl}.

The point-line geometry currently central to understanding some
particular aspects of the black hole analogy is the unique
generalized quadrangle of order $(2, 4)$, GQ$(2, 4)$ \cite{lsvp}.
Its Veldkamp space is isomorphic to PG$(5, 2)$ \cite{sglpv}, which
is the natural embedding space not only for GQ$(2, 4)$ itself, but
also for the split Cayley hexagon of order two and the Klein
quadric
--- other two prominent finite geometries linking quantum
information with black holes \cite{lsv}.  The geometry
GQ$(2, 4)$ contains
several notable subgeometries, one of them being the unique
generalized quadrangle of order two, GQ$(2, 2)$. This quadrangle
and its associated Veldkamp space ($\simeq$ PG$(4, 2)$) were found
to underlie the commutation relations between the elements of
two-qubit Pauli group and to also clarify some conceptual issues
of quantum mechanics \cite{spph,ps2,spp}.  Based on these
observations, Vrana and L\' evay \cite{vl}
have even discovered  a
whole family of Veldkamp spaces, which are associated with the
Pauli groups of qubits of arbitrary multiplicity.

To know the structure of a Veldkamp space is vital not only for
its own sake, but also because this space fully encodes
information about the point-line incidence geometry it is
associated with (or generated by).  In this paper, with our
interest mainly stirred by the above-described remarkable physical
applications, we shall focus on the Veldkamp space of the
point-line incidence structure that is
 known under several names: the smallest
slim dense near hexagon \cite{bchw,bruyn}, the $27_3$ Gray
configuration \cite{pis}, or, simply, a $(3 \times 3 \times 3)$-grid
(and in what follows denoted as $L_{3}^{\times 3}$). Although
$L_{3}^{\times 3}$ may seem to be a rather trivial geometry, its
Veldkamp space is endowed with a rich and complex structure well
worth deserving a closer look at.

Brouwer {\it et al.} \cite{bchw} proved that there are eleven isomorphism types
of slim dense near hexagons.  Of these eleven, the near hexagons of sizes $27$,
$45$ and $81$ are the most relevant for physical applications.  This paper
is devoted to a study of the first of the three examples; we plan to
describe the other two cases separately.  However, it is important to
emphasize at the very beginning that this paper is {\it not} about the
well-known universal embedding of $L_{3}^{\times 3}$ into PG(7,\,2).
Although it will be shown that our Veldkamp space is indeed isomorphic to
PG(7,\,2), the details of the isomorphism are nontrivial, and it is precisely
these details that are of interest for envisaged physical applications.

\section{Near Polygons, Quads, Geometric Hyperplanes and\\ Veldkamp Spaces}
In this section we gather all the basic  notions and well-established theoretical results that will be
needed in the sequel.

A {\it near polygon} (see, e.\,g., \cite{bruyn} and references
therein) is a connected partial linear space $S = (P, L, I)$, $I
\subset P \times L$, with the property that given a point $x$ and
a line $L$, there always exists a unique point on $L$ nearest to
$x$. (Here distances are measured in the point graph, or
collinearity graph of the geometry.)  If the maximal distance
between two points of $S$ is equal to $d$, then the near polygon
is called a near $2d$-gon. A near 0-gon is a point and a near
2-gon is a line; the class of near quadrangles coincides with the
class of generalized quadrangles.

A nonempty set $X$ of points in a near polygon $S = (P, L, I)$ is
called a subspace if every line meeting $X$ in at least two points
is completely contained in $X$. A subspace $X$ is called
geodetically closed if every point on a shortest path between two
points of $X$ is contained in $X$. Given a subspace $X$, one can
define a sub-geometry $S_X$ of $S$ by considering only those
points and lines of $S$ which are completely contained in $X$. If
$X$ is geodetically closed, then $S_X$ clearly is a
sub-near-polygon of $S$. If a geodetically closed sub-near-polygon
$S_X$ is a non-degenerate generalized quadrangle, then $X$ (and
often also $S_X$) is called a {\it quad}.

A near polygon is said to have order $(s, t)$ if every line is
incident with precisely $s+1$ points and if every point is on
precisely $t+1$ lines. If $s = t$, then the near polygon is said
to have order $s$. A near polygon is called {\it dense} if every
line is incident with at least three points and if every two
points at distance two have at least two common neighbours. A near
polygon is called {\it slim} if every line is incident with
precisely three points. It is well known (see, e.\,g.,
\cite{pay-thas}) that there are, up to isomorphism, three slim
non-degenerate generalized quadrangles. The $(3 \times 3)$-grid is
the unique generalized quadrangle of order $(2, 1)$, GQ$(2, 1)$.
The unique generalized quadrangle of order 2, GQ$(2, 2)$, is the
generalized quadrangle of the points and lines of PG(3, 2) which
are totally isotropic with respect to a given symplectic polarity.
The points and lines lying on a given nonsingular elliptic quadric
of PG$(5, 2)$ define the unique generalized quadrangle of order
$(2, 4)$, GQ$(2,4)$. Any {\it slim dense} near polygon contains
quads, which are necessarily isomorphic to either GQ$(2, 1)$,
GQ$(2, 2)$ or GQ$(2, 4)$.

Next, a {\it geometric hyperplane} of a partial linear space is a
proper subspace meeting each line (necessarily in a unique point
or the whole line). The set of points at non-maximal distance from
a given point $x$ of a dense near polygon $S$ is a hyperplane of
$S$, usually called the {\it singular} hyperplane with {\it
deepest} point $x$. Given a hyperplane $H$ of $S$, one defines the
{\it order} of any of its points as the number of lines through
the point which are fully contained in $H$; a point of a
hyperplane/sub-configuration is called {\it deep} if all the lines
passing through it are fully contained in the
hyperplane/sub-configuration. If $H$ is a hyperplane of a dense
near polygon $S$ and if $Q$ is a quad of $S$, then precisely one
of the following possibilities occurs: (1) $Q \subseteq H$; (2) $Q
\cap H = x^{\perp} \cap Q$ for some point $x$ of $Q$; (3) $Q \cap
H$ is a sub-quadrangle of $Q$; and (4) $Q \cap H$ is an ovoid of
$Q$. If case (1), case (2), case (3), or case (4) occurs, then $Q$
is called, respectively, {\it deep}, {\it singular}, {\it
sub-quadrangular}, or {\it ovoidal} with respect to $H$. If $S$ is
slim and $H_1$ and $H_2$ are its two distinct hyperplanes, then
the complement of symmetric difference of $H_1$ and $H_2$,
$\overline{H_1 \Delta H_2}$, is again a hyperplane; this means
that the totality of hyperplanes of a slim near polygon form a
vector space over the Galois field with two elements, GF(2). In
what follows, we shall put $\overline{H_1 \Delta H_2} \equiv H_1
\oplus H_2$ and call it the (Veldkamp) sum of the two hyperplanes.

Finally, we shall introduce the notion of the {\it Veldkamp space}
of a point-line incidence geometry $\Gamma(P, L)$,
$\mathcal{V}(\Gamma)$ \cite{buek}.  Here,  $\mathcal{V}(\Gamma)$  is the
space in  which (i) a point is a geometric hyperplane of  $\Gamma$
and (ii) a line is the collection $H'H''$ of all geometric
hyperplanes $H$ of $\Gamma$  such that $H' \cap H'' = H' \cap H =
H'' \cap H$ or $H = H', H''$, where $H'$ and $H''$ are distinct
points of $\mathcal{V}(\Gamma)$.
Following \cite{spph,sglpv}, we
adopt also here the definition of Veldkamp space given by
Buekenhout and Cohen \cite{buek} instead of that of Shult
\cite{shult}, as the latter is much too restrictive by requiring
any three distinct hyperplanes $H'$, $H''$ and $H'''$ of $\Gamma$
to satisfy the following two conditions: i) $H'$ is not properly contained
in $H''$ and ii) $H' \cap H'' \subseteq H'''$
implies $H' \subset H'''$ or  $H' \cap H'' = H' \cap H'''$. The two
definitions differ in the crucial fact that whereas the Veldkamp space in the sense
of Shult is {\it always} a linear space, that of Buekenhout and Cohen needs
not be so; in other words, Shult's Veldkamp lines are always
of the form $\{H \in \mathcal{V}(\Gamma) ~|~ H \supseteq H' \cap H''\}$ for certain geometric hyperplanes
$H'$ and $H''$.

\section{Veldkamp Space of $L_{3}^{\times 3}$}
\subsection{Geometric Hyperplanes}
Our point-line geometry $L_{3}^{\times 3}$ consists of 27 points
and the same number of lines, with three points on a line and,
dually, three lines through a point. Its full group of
automorphisms $G$ is isomorphic to $S_3 \wr S_3$, of order 1296
\cite{bchw}. As already mentioned, apart from being the smallest
slim dense near hexagon, it is also one of a pair of dual to each other
$27_3$ configurations
whose incidence
graph is the {\it Gray} graph \cite{pis}
--- the smallest cubic graph which is edge-transitive and regular,
but not vertex-transitive \cite{gray}. Being a slim geometry, its
geometric hyperplanes are readily found by employing the
above-mentioned property that the complement of symmetric
difference of any two of its hyperplanes is again a hyperplane.
There is a famous result of Ronan
\cite{ron} stating that if a point-line incidence geometry is
embeddable and is slim, then all of its geometric hyperplanes
arise from its universal embedding. As our geometry is indeed
universally embeddable into PG(7,\,2), we could readily get the
hyperplanes from the structure of this projective space. But as
the focus of our paper is on the community of physicists for which
the concept of Veldkamp space is virtually unknown, we opt below
for a less abstract and more elementary, diagrammatical exposition
of the topic.

\begin{figure}[pth!]
\centerline{\includegraphics[width=6cm,clip=]{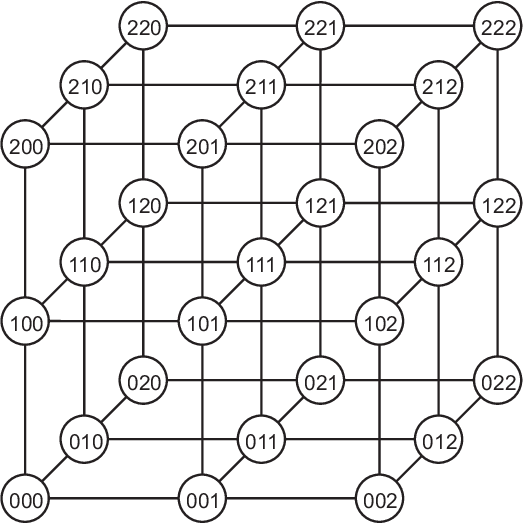}}
\vspace*{.2cm} \caption{A diagrammatic illustration of the
structure of $L_{3}^{\times 3}$. The points are represented by
circles and the lines by straight segments joining triples of
points. Also shown is a ternary digits labelling of the points,
employed in the sequel.}
\end{figure}

To this end, we shall use a diagrammatical representation of our
geometry as a $(3 \times 3 \times 3)$-grid, Figure 1. From this
representation we find that all its quads are GQ$(2,1)$s, and
there are altogether nine of them. Its singular hyperplane, $H_1$,
is endowed with 19 points and 15 lines, with 12 points of order
two and seven of order three, and three deep and six singular
quads
--- as easily discerned from Figure 2-1. To find the stabilizer
group of this type of geometric hyperplane,  one thinks of
$L_{3}^{\times 3}$ as a collection of triples of ternary digits
(Figure 1). We can act on the first, second and third digit
independently by permuting by $S_3$. Finally, we can permute the
digits themselves by another $S_3$. The stabilizer of an $H_1$ is
the same as the stabilizer of its deepest point, i.\,e. $Z_2 \wr
S_3$, of order 48; taking this point to be 000, the $Z_2$ decides
whether or not to exchange 1 and 2 in each position and the $S_3$
permutes the positions. There are altogether $|G|/|Z_2 \wr S_3| =
27$ copies of $H_1$ in $L_{3}^{\times 3}$.

Next, pick up two distinct copies $H_1(x)$ and $H_1(y)$ of $H_1$,
where $x$ and $y$ are the corresponding deepest points. If $x$ and
$y$ are at distance two from each other, then $H_1(x) \oplus
H_1(y) \equiv H_2$ is a new type of hyperplane, not isomorphic to
$H_1$. As depicted in Figure 2-2, an $H_2$ comprises 15 points and
nine lines, with six points of order one and two, and three deep
points; it features one deep, six singular and two ovoidal quads.
Its stabilizer is $Z_2 \times S_3 \times Z_2$, of order 24. The
first $Z_2$ switches the two planes/quads parallel to the included
one (0ab, say). We also need to permute the points 000, 011 and
022 among themselves; this means that we can do one permutation
($S_3$) affecting the second and third coordinates simultaneously,
and also swap the second and third coordinates (that is the second
$Z_2$). We find altogether $54$ copies of $H_2$ in $L_{3}^{\times
3}$.

\begin{figure}[pht!]
\centerline{\includegraphics[width=14.6cm,clip=]{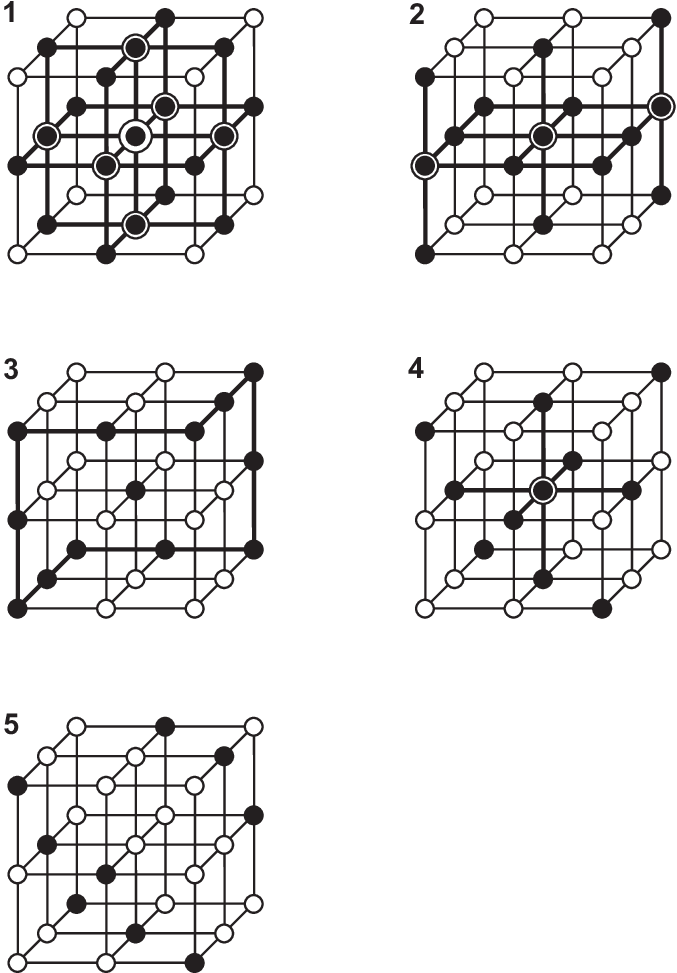}}
\vspace*{.2cm} \caption{A diagrammatic illustration of the five
distinct types of geometric hyperplanes of the smallest slim dense
near hexagon. The points/lines belonging to a hyperplane are
boldfaced; double circles stand for deep points.}
\end{figure}

If the two deepest points are at maximum distance from each other,
then $H_1(x) \oplus H_1(y) \equiv H_3$ is again a new kind of
hyperplane, not isomorphic to the previous two --- see Figure 2-3.
An $H_3$ comprises 13 points and six lines, with one point of
order zero (henceforth called the nucleus), six points of orders
one and two, and no deep points. There are no deep quads there,
only six singular and three ovoidal ones. The stabilizer is
$D_{12}$. For if one looks at the copy of $H_3$ shown in Figure
2-3 along the appropriate axis, it looks like a regular hexagon
with a dot (= the nucleus) in the middle; from this perspective,
the result is obvious. Altogether 108 copies of $H_3$ sit inside
$L_{3}^{\times 3}$.

There exist two more types of geometric hyperplanes. A hyperplane
of type $H_4$, Figure 2-4, features 11 points and three lines,
four points of order zero, six of order one and one deep point,
and, like the previous type, it exhibits only singular (three) and
ovoidal (six) quads. It is stabilized by $S_4$. This
can be seen
from Figure 2-4, as the four points of order zero form the
vertices of a regular tetrahedron and the symmetry group of our
$H_4$ is that of the tetrahedron (including reflections). One can
get a copy of $H_4$ in the following manner.  Take  a copy of
$H_2$; we already know that any $H_2$ is the complement of the
symmetric difference of two $H_1$s whose deeps are at distance two
from each other. An $H_4$ is then the complement of the symmetric
difference of the $H_2$ and the copy of $H_1$ whose deep is at
distance three from each of the previous two. Hence, one gets an
$H_4$ as the sum of {\it three} $H_1$s two of which deeps are at
distance two from each other and the third one being at the
maximum distance from the two. The cardinality of $H_4$ is 54.

%\newpage

The final type, $H_5$, is an ovoid of $L_{3}^{\times 3}$, that is
a set of nine pairwise non-collinear points (Figure 2-5). Hence,
in this case all the points are of zeroth order and all the quads
are ovoidal. The stabilizer group is $E \rtimes S_3$, where $E
\simeq (Z_3 \times Z_3) \rtimes Z_2$, and the $Z_2$ acts as inversion.
There is a nice description
of this using the ternary digits method. Take all coordinates
whose digits sum to zero mod 3. The  $S_3$ just permutes the
coordinates. The group $E$ has order 18 and is generated by an
$S_3$ and an element $g$ of order 3. The element $g$ acts by
increasing (respectively, decreasing) the second (respectively,
third) digit by 1 mod 3. The $S_3$ acts as the same permutation
(value permutation, not place permutation) on each digit
simultaneously. To get a copy of $H_5$, one selects an $H_3$, any
of which is the complement of the symmetric difference of two
$H_1$s whose deeps are at maximum distance from each other. A copy
of $H_5$ is then found as the complement of the symmetric
difference of the $H_3$ and the copy of $H_1$ whose deep is at
maximum distance  from either of the two. Hence, one gets an $H_5$
as the sum of {\it three} $H_1$s whose deeps are pairwise at
maximum distance from each other (compare with the preceding
case). There are only 12 copies of $H_5$ inside $L_{3}^{\times
3}$.

\begin{table}[t]
\begin{center}
\caption{The types of geometric hyperplanes of the smallest slim
dense near-hexagon.} \vspace*{0.3cm}
{\begin{tabular}{||l|c|c|c|c|c|c|c|c|c|c|c|r||} \hline \hline
\multicolumn{1}{||c|}{} & \multicolumn{1}{|c|}{} & \multicolumn{1}{|c|}{}  &  \multicolumn{4}{|c|}{} & \multicolumn{3}{|c|}{}  &\multicolumn{1}{|c|}{} &\multicolumn{1}{|c|}{}&\multicolumn{1}{|c||}{}\\
%\cline{4-11}
\multicolumn{1}{||c|}{} & \multicolumn{1}{|c|}{} &
\multicolumn{1}{|c|}{}  &  \multicolumn{4}{|c|}{$\#$ of Points of
Order} & \multicolumn{3}{|c|}{$\#$ of Quads of Type}
&\multicolumn{1}{|c|}{} &\multicolumn{1}{|c|}{}
&\multicolumn{1}{|c||}{}\\
 \cline{4-10}
Hp & Pts & Lns  & ~0~ & ~1~ & 2 & ~3~ & deep & ~sng~ & ovd & StGr & Wgt & Crd  \\
\hline
$H_1$ & 19 & 15  & 0 & 0 & ~12~ & 7 & 3 & 6 & 0 & $Z_2 \wr S_3$ & 1& 27  \\
$H_2$ & 15 & 9  & 0 & 6 & 6 & 3 & 1 & 6 & 2 &  $Z_{2}^{\times 2} \times S_3$ &2 & 54  \\
$H_3$ & 13 & 6  & 1 & 6 & 6 & 0 & 0 & 6 & 3 &  $D_{12}$ & 2& 108  \\
$H_4$ & 11 & 3  & 4 & 6 & 0 & 1 & 0 & 3 & 6 &  $S_4$ & 3& 54  \\
$H_5$ & 9 & 0  & 9 & 0 & 0 & 0 & 0 & 0 & 9 & $E \rtimes S_3$ &3 & 12  \\
\hline \hline
\end{tabular}}
\end{center}
\end{table}

The attentive reader might have noticed that $H_1$ plays a special
role amongst the hyperplanes, because each of the remaining types
can be obtained as the sum of several copies of $H_1$; the
smallest such number is called the {\it weight} of a hyperplane.
From what we said above it follows that $H_2$ and $H_3$ are of
weight two, whereas $H_4$ and $H_5$ have weight three. In this
respect, $L_{3}^{\times 3}$ resembles the {\it dual} of the split
Cayley hexagon of order two \cite{fj}, save for the fact that the
latter features many more types of geometric hyperplanes.

All our findings are summarized in Table 1. We see that the total
number of hyperplanes is $27 + 54 + 108 + 54 + 12 = 255$. As we
already know that hyperplanes from a GF(2)-vector space and $255 =
2^8 - 1 =$ $|$PG(7,2)$|$, one is immediately tempted to conjecture
that $\mathcal{V}(L_{3}^{\times 3})$ is isomorphic to PG$(7, 2)$.
In the next section we shall show that this is indeed so.\\

\subsection{Veldkamp Lines}

\begin{table}[pth!]
\begin{center}
\caption{The types of Veldkamp lines of the Veldkamp space of the
smallest slim dense near-hexagon. The details of the
symbols/notation are explained in the text.} \vspace*{0.6cm}
{\begin{tabular}{||r|l|l|c|c|c|c|c|r||} \hline \hline
\multicolumn{1}{||c|}{}  &  \multicolumn{2}{|c|}{} & \multicolumn{5}{|c|}{}  &  \multicolumn{1}{|c||}{}\\
%\cline{4-11}
\multicolumn{1}{||c|}{} &   \multicolumn{2}{|c|}{Core} & \multicolumn{5}{|c|}{Composition}  &\multicolumn{1}{|c||}{} \\
%\multicolumn{1}{||c|}{Hyperplane} & \multicolumn{1}{|c|}{Pts} & \multicolumn{1}{|c|}{Lns}  &  \multicolumn{1}{|c|}{0} & \multicolumn{1}{|c|}{1} & \multicolumn{2}{|c|}{0} & \multicolumn{1}{|c|}{3}
%& \multicolumn{1}{|c|}{deep} & \multicolumn{1}{|c|}{sing} & \multicolumn{1}{|c|}{ovoid} & \multicolumn{1}{|c|}{subq} &\multicolumn{1}{|c||}{} \\
 \cline{2-8}
Type & ~Pts & ~Lns  & $H_1$ & $H_2$ & $H_3$ & $H_4$ & $H_5$ & Crd  \\
\hline
1 &  ~15 & ~11  & 3 & -- & -- & -- & -- &  27 \\
2 & ~13 & ~8  & 2 & 1 & -- & -- & -- &  162 \\
3 & ~12 & ~6  & 2 & -- & 1 & -- & -- & 108  \\
4 &  ~11 & ~7  & 1 & 2 & -- & -- & -- &  81 \\
5 &  ~10 & ~4  & 1 & 1 & 1 & -- & -- & 648  \\
6 &  ~9 & ~6  & -- & 3 & -- & -- & -- & 18  \\
7 &  ~9 & ~4  & 1 & -- & 2 & -- & -- &  324 \\
8 &  ~$9_{(2)}$ & ~3c  & 1 & 1 & -- & 1 & -- &  324 \\
9 &  ~9 & ~3  & 1 & -- & 2 & -- & -- &  324 \\
10 &  ~9 & ~3p  & -- & 3 & -- & -- & -- &  18 \\
11 &  ~$9_{(3)}$ & ~3c  & -- & 3 & -- & -- & -- &  108 \\
12 &  ~8 & ~3  & -- & 2 & 1 & -- & -- & 648  \\
13 &  ~8 & ~2  & 1 & -- & 1 & 1 & -- &  648 \\
14 &  ~7 & ~3  & 1 & -- & -- & 2 & -- &  27 \\
15 &  ~7 & ~2p  & -- & 1 & 2 & -- & -- & 162  \\
16 & ~$7_{(2)}$  & ~2c   & -- & 1 & 2 & -- & -- & 324  \\
17 & ~$7_{(3)}$  & ~2c   & -- & 1 & 2 & -- & -- & 324  \\
18 &  ~$7_{[2]}$ & ~1  & -- & 2 & -- & 1 & -- & 162  \\
19 &  ~$7_{[1]}$ & ~1  & -- & 1 & 2 & -- & -- & 324  \\
20 &  ~7 & ~0  & 1 & -- & 1 & -- & 1 & 108  \\
21 &  ~7 & ~0  & 1 & -- & -- & 2 & -- &  108 \\
22 &  ~6 & ~2c  & -- & 1 & 1 & 1 & -- & 648  \\
23 &  ~6 & ~2p  & -- & -- & 3 & -- & --  & 108  \\
24 &  ~6 & ~1  & -- & -- & 3 & -- & -- &  648 \\
25&  ~$6_{[3]}$ & ~0  & 1 & -- & -- & 1 & 1 & 216  \\
26 &  ~$6_{[2]}$ & ~0  & -- & 2 & -- & -- & 1 & 108  \\
27 &  ~$6_{[1]}$ & ~0  & -- & 1 & 1 & 1 & -- & 648  \\
28 &  ~$6_{[0]}$ & ~0  & -- & -- & 3 & -- & -- & 36  \\
29 &  ~$5_{[1]}$ & ~1  & -- & 1 & -- & 2 & -- &  162 \\
30 &  ~$5_{[0]}$ & ~1  & -- & -- & 2 & 1 & -- & 648  \\
31 &  ~$5_{(2)}$ & ~0  & -- & 1 & 1 & -- & 1 &  324 \\
32 &  ~$5_{(1)}$ & ~0  & -- & 1 & -- & 2 & -- &  324 \\
33 &  ~$5_{(0)}$ & ~0  & -- & -- & 2 & 1 & -- & 648  \\
34 &  ~4 & ~0  & -- & -- & 2 & -- & 1 & 324  \\
35 &  ~$4_{(3:1)}$ & ~0  & -- & -- & 1 & 2 & -- & 216  \\
36 &  ~$4_{(2:2)}$ & ~0  & -- & -- & 1 & 2 & -- & 324  \\
37 &  ~3 & ~1  & -- & -- & -- & 3 & -- & 54  \\
38 &  ~$3_{[1]}$ & ~0  & -- & 1 & -- & -- & 2 &  54 \\
39 &  ~$3_{[0]}$ & ~0  & -- & -- & 1 & 1 & 1 & 216  \\
40 &  ~2 & ~0  & -- & -- & -- & 2 & 1 &  108 \\
41 &  ~0 & ~0  & -- & -- & -- & -- & 3 & 4  \\
 \hline \hline
\end{tabular}}
\end{center}
\end{table}

As each Veldkamp line is obviously of the form $\{H', H'', H'
\oplus H'' \}$, and since $L_{3}^{\times 3}$ is rather small,
diagrams of the from given in Figure 2 were easy to use to find
out ``by hand" all 41 different types of Veldkamp lines, whose
basic properties are listed in Table 2. A rather coarse
classification in terms of the composition is refined by that of
the properties of the cores. To classify each type unambiguously,
we have to further refine, in the relevant cases, the core line
cardinality entries by stating whether the lines of the core are
concurrent (``c'') or parallel/mutually skew (``p'') and specify
the core point cardinality entries even in a greater variety of
ways, as follows:

\begin{itemize}

\item $4_{(3:1)}$ means that one of the points is at maximum distance from
the other three, in contrast to $4_{(2:2)}$ where two points are
such that either of them is at maximum distance from the remaining
three (to distinguish type 35 from type 36);

\item $7_{(2/3)}$ or $9_{(2/3)}$ stand for the fact that the two isolated
(= zeroth order) points of the core are at distance 2/3 from each
other (to distinguish type 16 from type 17);

\item $X_{[n]}$ illustrates the fact that there exist  exactly $n$ quads
each of which cuts the core in an ovoid, i.\,e. a triple of
pairwise non-collinear points (to distinguish types 25 to 28 from
one another); and, finally,

\item $5_{(n)}$ tells us that out of five points there are $n$ such that
each is coplanar (that is, shares a quad) with any of the
remaining four.

\end{itemize}

\noindent Table 2 reveals a number of interesting properties of
the Veldkamp lines. First, we notice that there is only one type
(41) whose core is an empty set; all the three Veldkamp points of
this line are of the same kind, namely $H_5$. Next, there are nine
different types of Veldkamp lines each of which consists of the
hyperplanes of the same type; these are types 1 ($H_1$), 6, 10, 11
($H_2$), 23, 24, 28 ($H_3$), 37 ($H_4$) and 41 ($H_5$). On the
other hand, we find nine (out of theoretically possible 10 = ${{5
\choose 3}}$) types featuring three different kinds of
hyperplanes; these are types 5, 8, 13, 20, 22, 25, 27, 31 and 39.
Interestingly, the only nonexistent
 ``heterogeneous'' combination
is the $H_1$-$H_2$-$H_5$ one. The most ``abundant" type of
Veldkamp points is $H_3$, which occurs in 23 types of lines,
followed by $H_2$ (20), $H_4$ (17), $H_1$ (13)
and finally
by $H_5$ (9). Remarkably, $H_3$ also prevails in multiplicity
$\geq 2$ (12 types), whereas in ``singles'' the primacy belongs to
$H_2$ (13 types). It is also worth mentioning that there are as
many as four different types of the form $H_2$-$H_3$-$H_3$ (15,
16, 17 and 19). All these observations are gathered in Table 3.

\begin{table}[t]
\begin{center}
\caption{A succinct overview of the composition of the Veldkamp
lines; here, for example, number 6 in the row $H_2$ and the column
$H_4$ means that there are 6 different types of lines featuring
both $H_2$ and $H_4$. } \vspace*{0.6cm}
\begin{tabular}{|c|ccccc|}
\hline \hline
  & ~$H_1$ & $H_2$ & $H_3$ & $H_4$ & $H_5$~  \\
\hline
~$H_1$~ &    3  &   4   &    6  &   5   & 2 \\
$H_2$ &      &   7   &    9  &   6   & 3 \\
$H_3$ &      &      &    12  &   8   & 4 \\
$H_4$ &      &      &      &   8   & 3 \\
$H_5$ &      &      &      &      & 2 \\
 \hline \hline
\end{tabular}
\end{center}
\end{table}

The structure of Table 3 can easily be demystified if one looks at
a decomposition of the group of automorphisms into double cosets
with respect to stabilizer (sub-)groups of hyperplanes. Let us
recall (see, e.\,g., \cite{bog}) that given a group $G$ and its
two (not necessarily distinct) subgroups $K_1$ and $K_2$, the set
$$K_1 a K_2 \equiv \{k_1 a k_2 \in G | a \in G, k_1 \in K_1, k_2
\in K_2 \}$$ is called a double coset with respect to $K_1$ and
$K_2$. So, for $G \simeq S_3 \wr S_3$ and with $K_i$, $i=1, 2$,
identified with the stabilizers of five different kinds of
hyperplanes, one finds
--- using, for example, a GAP code \cite{gap} --- the number of representatives of double cosets
for different combinations of hyperplanes as given in Table 4.
Comparing the last two tables one sees that they are {\it almost}
identical; the only discrepancies occur in the $H_3$-$H_3$,
$H_3$-$H_4$ and $H_4$-$H_4$ entries, where we always find more
double coset representatives than line types. This means that
there are  particular Veldkamp line types that  correspond to more
than one double coset. And, indeed, after some work one finds that
types 33 and 35 correspond each to two double coset
representatives, whereas type 24 corresponds to as many
as  three. To
illustrate the origin of this puzzling ``multi-valuedness''
property, we shall consider the last mentioned case.

\begin{table}[t]
\begin{center}
\caption{The set of representatives, after discarding the relation
of equality, of double cosets of the group $G$ of automorphisms of
the near hexagon with respect to the stabilizer groups of all five
different kinds of hyperplanes. By abuse of notation, the
stabilizer group of a hyperplane is denoted by the same symbol as
the hyperplane itself. Thus, for example, the entry 6 in the row
$H_2$ and the column $H_4$ should read that there are 6 distinct
double cosets $H_2 a H_4$, $a \in G$.} \vspace*{0.6cm}
\begin{tabular}{|c|ccccc|}
\hline \hline
  & ~$H_1$ & $H_2$ & $H_3$ & $H_4$ & $H_5$~  \\
\hline
~$H_1$~ &    3  &   4   &    6  &   5   & 2 \\
$H_2$ &      &   7   &    9  &   6   & 3 \\
$H_3$ &      &      &    15  &   9   & 4 \\
$H_4$ &      &      &      &   9   & 3 \\
$H_5$ &      &      &      &      & 2 \\
 \hline \hline
\end{tabular}
\end{center}
\end{table}

We first introduce some more terminology. Given a copy of $H_3$
and disregarding its nucleus, the 12 remaining points are such
that there exist exactly two distinct points coplanar with all of
them. Let us call this pair of points the ``axis'' of the $H_3$.
Next, let us select three distinct copies $A$, $B$, and $C$ of
$H_3$ as follows (employing again our ternary digits
representation of the points of the near hexagon): $A$ is the
$H_3$ with nucleus 111 and axis $\{000, 222\}$, $B$ is the $H_3$
with nucleus 020 and axis $\{102, 211\}$, and $C$ is the $H_3$
with nucleus 122 and axis $\{001, 210\}$. It is not difficult to
verify that the triple $\{A, B, C\}$, with its core comprising the
points 012, 020, 111, 200, 201 and 202, is indeed a Veldkamp line
of type 24. Now, consider the six {\it ordered} pairs $\{(A, B),
(A, C), (B, A), (B, C), (C, A), (C, B) \}$.  The orbits of $G$
acting on this set are $\{(A, B), (B, A)\}$,  $\{(A, C), (B,
C)\}$, and $\{(C, A), (C, B)\}$ because $(A, C)$ is not
$G$-conjugate to $(C, A)$ even though $A$ is $G$-conjugate to $C$!
This just means that  the
{\it single} type of a Veldkamp line,
that is to say the triple $\{A, B, C\}$,  corresponds to {\it
three} distinct double cosets; one of them is symmetric, the other
two are  transposes of each other.

The remaining issue is to find the cardinality for each type of Veldkamp
lines and verify that their sum indeed amounts to 10\,795, that is to the number of lines
of PG$(7, 2)$.
These calculations were mostly done by computer, again
using  a particular GAP code \cite{gap},
and their results are given in the last column of Table 2.
It must be stressed that types 6, 23, 37 and 41 behave in a rather peculiar way as in these cases it is impossible
to fully reconstruct/recover the hyperplanes of the line from the configuration of the core and, therefore, one has to
be more careful in counting here. It is also worth noting that there are only four distinct Veldkamp lines
with the empty core, that cardinalities 4, 36 and 81 occur just for a single line type, and that there exist as many as 10
different types of lines of cardinality 324; the next most frequent cardinality
is 648, occurring in 8 distinct types.

We shall conclude this section by the following important observation. Suppose we have a Veldkamp space
over GF$(2)$ whose associated point-line incidence geometry has an odd number of points and
also all its geometric hyperplanes have an odd cardinality. Then, given any two distinct hyperplanes $H'$ and $H''$, there exists on this Veldkamp space a $G$-invariant GF(2)-bilinear form
$B(H', H'')$ defined as follows: $B(H', H'') = 0$ if $|H' \cap H''|$ is odd and $B(H', H'') = 1$ if
$|H' \cap H''|$  is even.
The $G$-invariance of the form is immediate, and it turns out that
this is the only
nontrivial $G$-invariant bilinear form over GF(2) on the Veldkamp space.
The GF(2)-bilinearity stems from the facts that: a) if $H'$, $H''$ and
$H'''$ form a Veldkamp line then their  pairwise intersection is the same as their triple intersection and b) the total number of points is
odd and so is is the number of points in each hyperplane. In our case, the corresponding form is the unique symplectic form
of PG$(7, 2)$ with respect to which the totality of 10\,795 Veldkamp lines split into two disjoint sets: 5\,355 isotropic lines (odd core point cardinality) and
5\,440 non-isotropic ones (even core point cardinality).

\section{Summary and Conclusion}
A comprehensive description of the Veldkamp space of the smallest
slim dense near hexagon ({\it alias} the Gray configuration, or a
$(3 \times 3 \times 3)$-grid ) has been presented. Being isomorphic
to PG$(7, 2)$, its $2^8 - 1 = 255$ Veldkamp points fall into five
distinct classes. Each class is uniquely characterized by the
number of points/lines as well as by a sequence of the
cardinalities of points of given orders and/or that of
(grid-)quads of given types; in addition, we also provide its
weight, stabilizer group within the full automorphism group of the
near hexagon and the total number of copies. Similarly, the
totality of ${{255 \choose 2}}$/3 = 10\,795 Veldkamp lines are
shown to form 41 different types. A complete classification of
them, partly based on the properties of double cosets of the group
of automorphisms with respect to the stabilizer groups of
hyperplanes, is given in terms of the properties of their cores
and the types of the hyperplanes they are composed of. Given a specific symplectic polarity in
PG$(7, 2)$, we also show that isotropic/non-isotropic Veldkamp lines are
of those types whose core point cardinality is odd/even.
It is establishing the properties of the Veldkamp lines and their cores, some of which
are quite intricate and were rather difficult to discern, which we regard
as the main result of the paper. In a similar way to what the Veldkamp lines of GQ(2,\,2) revealed to us about the finer structure of the algebra
of a two-qubit Pauli group \cite{spph}, knowing the structure of the cores of our near hexagon will be
vital for getting deeper insights into the commutation algebra of quantum operators/observables
underlined by this incidence geometry.

As already stressed, we expect this space to have interesting
physical applications. We just give an outline of a couple of
arguments giving support to such expectations.
It is a well-known
fact that the group $Spin(8)$, the double-cover of $SO(8)$, has
exactly three irreducible real representations of degree eight
and, accordingly, three representation spaces: one vectorial and
two spinorial. Remarkably, the three spaces are on equal footing
and there exits an extra automorphism, known as {\it triality},
that permutes them. This feature has already been found to play a
distinguished role in certain supersymmetry and supergravity
theories (see, e.\,g., \cite{adams}). Strikingly, it is also our
Veldkamp space that is endowed with a {\it triality}, albeit a
geometric one.
To be more specific (see, e.\,g., \cite{ht}), let us consider a
non-singular hyperbolic quadric $Q^{+}(7, 2)$ in PG$(7, 2)$. The
generators of this particular quadric are three-dimensional
subspaces. The set of generators can be divided into two families
$P_1$ and $P_2$ such that two distinct generators belong to the
same family if and only if they intersect in a subspace of odd
dimension (i.\,e., in the empty set, a line or a three-dimensional
subspace (in which case they coincide)). Let $P_0$ be the set of
points of $Q^{+}(7, 2)$ and call the elements of $P_i$ the
$i$-points, $i \in \{0, 1, 2\}$; let $L$ denote the set of lines
on $Q^{+}(7, 2)$. A simple counting argument yields $|P_0| = |P_1|
= |P_2|$. A {\it triality} of $Q^{+}(7, 2)$ is a bijection $\mu$
that maps: $P_0 \mapsto P_1$, $P_1 \mapsto P_2$, $P_2 \mapsto
P_0$, $L \mapsto L$, is incidence preserving  and satisfies
$\mu^{3} = 1$. An absolute $i$-point is an $i$-point that is
incident with its image under the triality $\mu$, $i \in \{0, 1,
2\}$; an absolute line is a line that is fixed by $\mu$. For
certain trialities the incidence structure formed by the absolute
$i$-points and the absolute lines, with given incidence, is a
generalized hexagon. And for a {\it specific} triality we get the
{\it split Cayley hexagon of order two} --- an important
representative of finite geometry found to underlie the properties
of entropy formulas of a distinguished class of stringy black
holes \cite{lsv}.

The second argument is related to the concept of superqubits as proposed in
a very recent paper by Borsten {\it et al.} \cite{brst}. Here, the structure
of $L_{3}^{\times 3}$ seems to be {\it directly} related to the Hilbert space
of three-superqubits. This super Hilbert space is 27-dimensional and associating
bijectively its dimensions with the  27 points of $L_{3}^{\times 3}$ one immediately
finds that its 13 fermionic dimensions should correspond to a geometric hyperplane
of type $H_3$ and its 14 bosonic ones to the complement of the hyperplane in question.
Moreover, any truncation to two-superqubits would correspond to a GQ$(2, 1)$-quad in
$L_{3}^{\times 3}$, and the associated splitting of the 9-dimensional super Hilbert
sub-space into 5 boson and 4 fermion dimensions would simply correspond
to a perp-set (of the GQ$(2, 1)$) and its complement, respectively.
A firmer
footing for this argument can only be given after we are familiar with the structure
of the Veldkamp spaces of other two closely-related point-line incidence geometries, namely of L$_3 \times$ GQ(2,\,2)
and L$_3 \times$ GQ(2,\,4), in each of which our near hexagon sits as a subgeometry.
We are almost done with the first case \cite{sng}, but still far from being so for the second geometry.

Further explorations along these lines are clearly vital as they may, among other
things, shed further light on the relation between quantum
information theory and physics of stringy black holes outlined in
the introduction. In view of these facts, proper understanding of
the structure of $\mathcal{V}(L_{3}^{\times 3})$ --- as expounded
in preceding sections --- is of great importance.

\vspace*{.1cm} \noindent
{\bf Acknowledgements}\\
\normalsize M. S. was partially supported by the VEGA grant agency
projects Nos. 2/0092/09 and 2/0098/10, as well as by the ZiF 2009
Cooperation Group Project ``Finite Projective Ring Geometries: An
Intriguing Emerging Link Between Quantum Information Theory,
Black-Hole Physics and Chemistry of Coupling'' (Center for
Interdisciplinary Research (ZiF), University of Bielefeld,
Bielefeld, Germany).
We are grateful to Petr Pracna for
computer versions of the figures.

\end{document}